\documentclass[prl,twocolumn,preprintnumbers,amsmath,amssymb]{revtex4-1}
\usepackage[english]{babel}
\usepackage{amsmath,amsfonts,amssymb,hyperref}
\usepackage{graphicx}
\usepackage{dcolumn}
\usepackage{bm}
\usepackage{color}

\newcommand{\df}{\text{d}}

\begin{document}
\renewcommand{\figurename}{FIG.}
\title{Quantum radiation reaction in laser-electron beam collisions}
\author{T. G. Blackburn$^1$, C. P. Ridgers$^{2,3}$, J. G. Kirk$^4$, A. R. Bell$^{1,3}$}

\address{{\small{\small
{$^1$Clarendon Laboratory, University of Oxford, Parks Road, Oxford, OX1 3PU, UK \\
$^2$Department of Physics, University of York, York, YO10 5DD, UK \\
$^3$Central Laser Facility, STFC Rutherford-Appleton Laboratory, Chilton, Didcot, Oxfordshire, OX11 0QX, UK \\
$^4$Max-Planck-Institut f\"{u}r Kernphysik, Postfach 10 39 80, 69029 Heidelberg, Germany}
}}}

\begin{abstract}
It is possible using current high intensity laser facilities to reach the
quantum radiation reaction regime for energetic electrons. An experiment
using a wakefield accelerator to drive GeV electrons into a 
counterpropagating laser pulse would demonstrate the increase in the yield
of high energy photons caused by the stochastic nature of quantum
synchrotron emission: we show that a beam of $10^9$ 1 GeV electrons
colliding with a 30 fs laser pulse of intensity $10^{22} \, \text{Wcm}^{-2}$
will emit $6300$ photons with energy greater than 700 MeV, $60\times$ the
number predicted by classical theory.
\end{abstract}

\maketitle

	\begin{figure}
	\includegraphics[width=\linewidth]{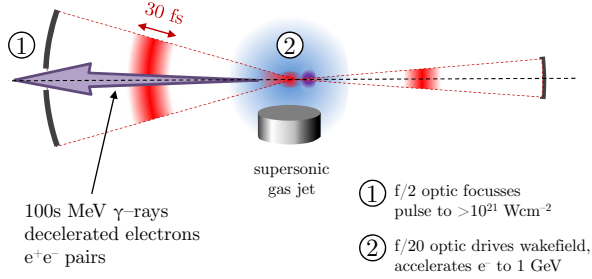}
	\caption{(Color online). Diagram of an experimental geometry that could
			demonstrate quantum radiation reaction and pair production.
			The GeV electrons decelerate in large fields at the laser
			focus, producing gamma rays that pass through a hole in the
			f/2 optic.}
	\label{fig:diagram}
	\end{figure}

Now that the power focussed in laser facilities exceeds 1 PW, electron
dynamics enters a regime between classical and QED physics. In this letter we
will consider an experimental setup where this transition can be explored:
the collision of a GeV electron beam and a high-intensity laser pulse
operating close to the current intensity frontier ($10^{22}\,\text{Wcm}^{-2}$),
as shown in Fig. \ref{fig:diagram}.
Electron dynamics in this regime is dominated by radiation
reaction \cite{DiP, DiP2}, which will manifest itself in the efficient conversion
of electron energy to hard gamma rays. We show that with current high-intensity
laser facilities, it is possible to provide the first demonstration of the
probabilistic nature of radiation reaction in the multi-photon, strong-field
QED regime. The stochastic nature of this process increases its efficiency by
over an order of magnitude compared to classical radiation reaction, and provides
the most sensitive diagnostic of quantum synchrotron emission.

This experimental setup will provide a means of testing the fundamental
physics underlying more exotic phenomena, such as pair cascades
\cite{Elkina, Ner}, which occur in pulsar magnetospheres \cite{Ast, Tim} and
are predicted to become significant above a threshold between $10^{23}$ and
$10^{24} \, \text{Wcm}^{-2}$. In these cases the plasma dynamics is strongly
affected by the generation of macroscopic electron-positron pair plasmas, for
example at the laser focus in laser-solid target interactions \cite{Ridgers}.

We parameterise the importance of strong-field QED effects with the quantity
$\eta = |F_{\mu\nu}p^\nu|/mcE_\text{Sch} \simeq \gamma |\mathbf{E}_\perp +
\mathbf{v} \times \mathbf{B}|/E_\text{Sch}$ \cite{Bell}, for an electron with
four-momentum $p^\mu = \gamma m (c, \mathbf{v})$ in an electromagnetic field
$F_{\mu\nu}$ with magnetic component $\mathbf{B}$ and component of the electric
field perpendicular to the electron direction of propagation $\mathbf{E}_\perp$.
$\eta$ is the ratio
of the electric field in the electron rest frame to $E_\text{Sch} =
m^2c^3/e\hbar$, the characteristic field of QED \cite{Sch}.
In the setup we consider, the probabilistic nature of photon
emission is evident for $\eta \sim 0.1$. The QED-dominated regime is reached
when $\eta > 1$; at $\eta = 90$, a 50 GeV electron would be capable of
producing multiple pairs on passage through a petawatt laser pulse
\cite{Sok}.

In a classical description, the electron travels on its worldline radiating
continuously as it accelerates in the laser fields. The typical
energy of a single emitted photon is $0.44 \eta \gamma mc^2$ \cite{Bell};
therefore as $\eta$ approaches 1, a single photon becomes capable of carrying off
a significant fraction of the electron's energy and the recoil of that emission
must be taken into account. However, the quantum description of radiation reaction
differs from classical theory in two ways.

Firstly, quantum corrections to the radiated spectrum cut off the tail of photons
with energies greater than that of the electron \cite{Ritus} and include spin-flip
transitions, the latter markedly increasing the probability of radiating photons with
energies comparable to that of the electron \cite{Ugg2}. These modifications means that
the total radiated power is smaller than the equivalent classical power by a factor
$g(\eta) \in$ (0,1) \cite{Erber, Baier}.

The second and more significant effect for us is that the process of photon
emission is stochastic \cite{Ner, Rol}; thus the electron has only a probability to
emit a gamma-ray photon of given energy. This gives rise to a phenomenon called
`straggling' \cite{Rol, Shen}, where the electron may propagate a significant
distance through the strong laser fields without radiating.
Since the laser pulse will have a spatial intensity profile, it is possible
for the electron to reach the region of highest intensity at the centre
having lost much less energy than a classical electron. The $\eta$ of an
electron that has straggled in this way will be boosted above that which
could be reached classically. As the tail of the photon spectrum increases non-linearly
with $\eta$, straggling enhances the yield of hard gamma-rays. Furthermore, these energetic photons 
can decay in the strong electromagnetic fields to produce electron-positron pairs
by the Breit-Wheeler process \cite{Bell}. Pairs can also be produced directly
by the electron in the trident process.

These strong-field QED effects have been investigated experimentally with 100 GeV
electrons incident on crystals at the CERN Super Proton Synchrotron \cite{Ugg} and
in the collision of a 50 GeV electron beam and a $10^{18}\,\text{Wcm}^{-2}$ laser
pulse at the SLAC facility \cite{SLAC}. In the former, the reduction in radiated
power $g(\eta)$ was measured and found to agree well with theoretical predictions;
in the latter, pair production by gamma rays created by inverse Compton
scattering was observed. However, by using a more intense laser, we can reach the
same $\eta$ with electrons of much lower energy. As it has been shown that PW lasers
in long focus can generate GeV electron beams by wakefield acceleration \cite{Kneip,
Kneip2, Kim}, it is realistic to consider the experiment shown in Fig.
\ref{fig:diagram}, where another laser pulse, in tight focus, provides the high-intensity target for the wakefield-accelerated electron beam.

We have developed a Monte-Carlo algorithm to simulate the interaction of an intense
laser pulse with an energetic electron beam, following \S2 of \cite{Rol} and \S3 of
\cite{JCP}. The three processes included are:
photon and (trident) pair production by an electron, and (Breit-Wheeler) pair
production by a photon.

The rates of these three processes are calculated in the Furry
picture of QED \cite{Furry}, where the electron interacts with both an external,
unquantised electromagnetic field and a fluctuating component of the same
\cite{Brown, Heinzl}. In between interactions with the latter (i.e. photon emission
and absorption) it propagates classically \cite{Mack2}. The photon formation
length is smaller than the laser wavelength by a factor of the laser strength
parameter $a_0$ \cite{Arka}; since $a_0 = [I_\text{L}
(\lambda/\mu \text{m})^2 /1.37 \times 10^{18} \, \text{Wcm}^{-2}]^{1/2} \gg 1$,
where $I_\text{L}$ and $\lambda$ are the laser intensity and wavelength
respectively, we treat the emission process as pointlike and instantaneous.
As the fields are quasi-static over the emission process, these rates may
be calculated in an equivalent system of fields with the same instantaneous value of $\eta$, such
as a static magnetic field (in the limit $B \rightarrow 0$, $\gamma \rightarrow \infty$)
or a plane EM wave (in the limit of zero frequency) \cite{Arka}.

Using the static magnetic field approach, the spectrum of emitted photons
is determined by the quantum synchrotron function $F(\eta, \chi)$, where
$\chi = (\hbar/2mc^2) |\omega \mathbf{E}_\perp + c^2\mathbf{k} \times 
\mathbf{B}|/E_\text{Sch}$ is the counterpart of the electron parameter
$\eta$ for a photon of frequency $\omega$ and wavevector $\mathbf{k}$.
The rate of emission is given in Erber \cite{Erber} and Baier {\it et al.}
\cite{Baier}. In the plane wave approach, this same process is called
non-linear inverse Compton scattering \cite{Mack, Mack2, Seipt}.

We further assume that the electrons may be treated independently as they are highly energetic,
with $\gamma \gg a_0 \gg 1$. The rigid-beam approximation is valid as the
transverse momentum gained from the laser is small, and in the frame
co-moving with a nearly mono-energetic beam, the energy of interaction
between particles will also be small. Therefore the electron motion, initially
antiparallel to the optical axis, is one-dimensional and constant between discrete
photon emission events.

This algorithm is implemented thus: the code initially assigns to each electron a pseudorandom `final' optical
depth against photon emission $\tau_\text{ph}$ and trident pair production
$\tau_\text{tri}$. As they propagate through the laser pulse, it integrates
their differential optical depths against trident pair production 
$\df\tau_\text{tri}/\df t$ and photon emission
$\df\tau_\text{ph}/\df t = (\sqrt{3}\alpha/2\pi\tau_\text{C}) \eta h(\eta)/\gamma$,
where $\alpha$ is the fine structure constant, $\tau_\text{C}$ the
Compton time and $h(\eta) = \int_0^{\eta/2}\!\df\chi F(\eta,\chi)/\chi$
\cite{Erber, Baier}.

Emission occurs when the electron reaches that `final' optical depth \cite{Rol}.
Following that, a new final depth is assigned. The photon energy is found by
pseudorandomly sampling the quantum synchrotron distribution \cite{Elkina, Sok}
$(\int_0^{\chi}\!\df\chi F(\eta,\chi)/\chi)/h(\eta)$.
The electron energy, constant between emissions, is then reduced by the energy
of the emitted photon. In the static magnetic field approach, this follows
from energy conservation. Since synchrotron radiation does not conserve the
component of momentum normal to the magnetic field \cite{LanLif}, when
implemented in the code, this leads to a fractional error
$\Delta\gamma/\gamma \propto 1/\gamma_\text{emit}$ which is negligible for
all emission events \cite{Ner, Ridgers}.

Energy loss to trident pair production is neglected, as the rate is calculated
in the Weizs\"{a}cker-Williams approximation, which treats the exchanged
virtual photon as real (i.e. the momentum transfer from electron to pair
$q^2 \rightarrow 0$). Breit-Wheeler pair production is modelled by assigning
to each photon at its creation a final optical depth $\tau_\text{BW}$.
The rate of pair production \cite{Erber} is integrated along the photon
trajectory to determine whether it decays to a pair.

We will compare our fully stochastic model of radiation reaction with a
semi-classical model, which we call `continuous radiation reaction'. In
this model the electron loses energy continuously according to a damping term
given by the Landau-Lifshitz force \cite{LanLif} modified to include the
quantum correction to the radiated power:
$\df\gamma/\df t = (2 \alpha/3) (\eta^2/\tau_\text{C})g(\eta)$.
For consistency,
photon spectra are obtained by sampling at each timestep the quantum synchrotron
distribution. Thomas {\it et al.} \cite{Thomas} use a similar model to simulate
laser-electron collisions for electrons with $\gamma_0 = 400$, where $\eta$ is
small enough that probabilistic effects can be neglected.

	\begin{figure}
	\includegraphics[width=\linewidth]{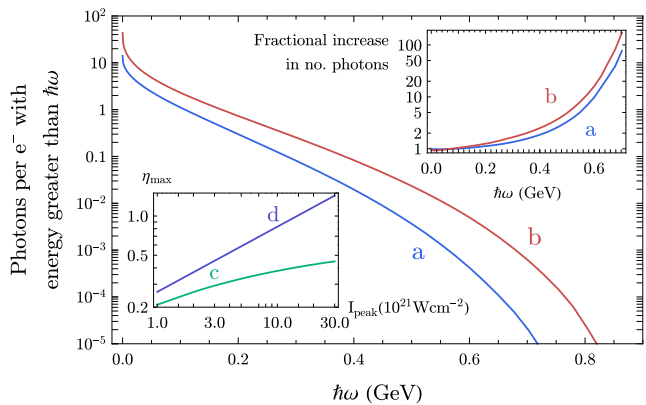}
	\caption{(Color online). The number of photons with energy $> \hbar
			\omega$ for a 1 GeV electron incident on a linearly polarised,
			plane-wave
			laser pulse with Gaussian temporal profile (FWHM 30 fs) and
			peak intensity (a) $10^{21}$ (b) $10^{22} \, \text{Wcm}^{-2}$.
			{\it Inset} ({\it top}) The fractional increase, due to straggling,
			in the number of photons with given minimum energy.
			{\it Inset} ({\it bottom}) The maximum $\eta$ experienced by
			a (c) continuously and (d) discontinuously radiating
			GeV electron incident on the same pulse with given intensity.
			The increase in going from (c) to (d) is responsible for
			the hardening of the photon spectrum.}
	\label{fig:photon1}
	\end{figure}

The parameters of the simulation are as follows: the laser pulse has wavelength
$\lambda = 1 \, \mu$m, is linearly polarised and has Gaussian temporal profile
with a full width at half-maximum (FWHM) of 30 fs. The electrons have initial
gamma factor $\gamma_0 = 2 \times 10^3$ and propagate along the optical axis
antiparallel to the laser pulse; this ensures that the $\mathbf{E}_\perp$ and
$\mathbf{v} \times \mathbf{B}$ terms in the definition of $\eta$ are additive,
i.e. $\eta \approx 2 \gamma E_\text{L}/E_\text{Sch}$. Each simulation follows
at least $10^7$ macroelectrons, their trajectories discretised into intervals of
$\Delta t = 10^{-17}$s. The laser fields are
calculated at each timestep using the classical solution for a Gaussian focussed
beam.

If the laser pulse is a plane wave, the number of photons per electron with
energy greater than $\hbar\omega$, and the increase in the same due to straggling,
are given in Fig. \ref{fig:photon1}. The boost to the electron's maximum $\eta$
as a result of straggling is also shown: this boost is responsible for the
hardening of the photon spectrum.

In reality both the laser and electron beam will have temporal and spatial
structure. As long as enough of the beam is incident on the laser focus,
there will still be strong evidence of probabilistic photon emission.
Consider a beam of 1 GeV electrons propagating antiparallel to the laser
but uniformly distributed around the optical axis in a disk of radius 10
$\mu$m. The laser pulse, still linearly polarised, is focussed with Gaussian
temporal and radial profiles (FWHM 30 fs and waist size 2.2 $\mu$m). If it
has peak intensity of $10^{22} \, \text{Wcm}^{-2}$, the number of photons per
electron with energy greater than 500 (700) MeV this interaction produces
will be $7.40\times10^{-4}$ ($1.32\times10^{-5}$), $4.3\times$ ($160\times$)
greater than that which would be radiated classically.

	\begin{figure}
	\includegraphics[width=\linewidth]{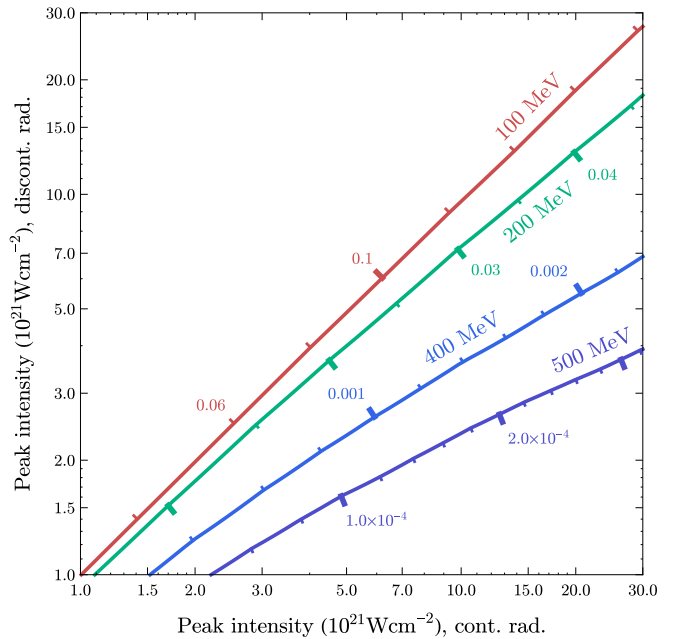}
	\caption{(Color online). Indicated along the coloured lines are the numbers
			of photons per electron with energy greater than 100 MeV (red),
			200 MeV (green), 400 MeV (blue) and 500 MeV (purple) if the
			electrons 	radiate discontinuously (vertical axis) or continuously
			(horizontal axis). The interaction here is between a focussed laser
			pulse of given peak intensity (FWHM 30 fs, waist 2.2 $\mu$m)
			and a beam of 1 GeV electrons (radius 10 $\mu$m around the
			optical axis).}
	\label{fig:photon2}
	\end{figure}

It is then possible to distinguish discontinuous radiation reaction by
measuring the high-energy tail of the gamma-ray spectrum, as the yield of
these photons is most enhanced by straggling. Fig. \ref{fig:photon2} shows
that if the electrons radiate semi-classically, rather than stochastically,
the laser pulse must be much more intense for the interaction to produce
the same number of high-energy photons. This increase in intensity is
necessary because a stochastically radiating electrons can reach a higher $\eta$.

Assessing the nature of the emitted radiation in a laser-electron beam
experiment can then done by fitting the observed spectrum to the points in
Fig. \ref{fig:photon2}. For example, the detection of $9.0\times10^{-2}$
photons per electron with energy greater than 100 MeV would be consistent
with either a continuously or discontinuously radiating beam colliding
with a laser pulse of peak intensity $5 \times 10^{21} \, \text{Wcm}^{-2}$.
In contrast, the simultaneous detection of $1.9\times10^{-3}$
photons per electron with energy greater than 400 MeV
would only be consistent with discontinuous radiation reaction,
as a continuously radiating beam would have to collide with a pulse of peak
intensity $1.8 \times 10^{22} \, \text{Wcm}^{-2}$ that number of gamma rays.
If the number of photons with energy greater than 500 MeV is measured instead,
that equivalent intensity rises to $5.2\times10^{22}\,\text{Wcm}^{-2}$.

A typical laser wakefield will accelerate a bunch of $10^9$ electrons, which
will produce a beam of $4.1\times10^5$ photons with $\hbar \omega > 500$
MeV, collimated within an angle of 30 mrad. When shielded to block lower
energy gamma rays, this will be sufficient energy deposition in a calorimeter
to provide a good signal of discontinuous radiation reaction. 

	\begin{figure}
	\includegraphics[width=\linewidth]{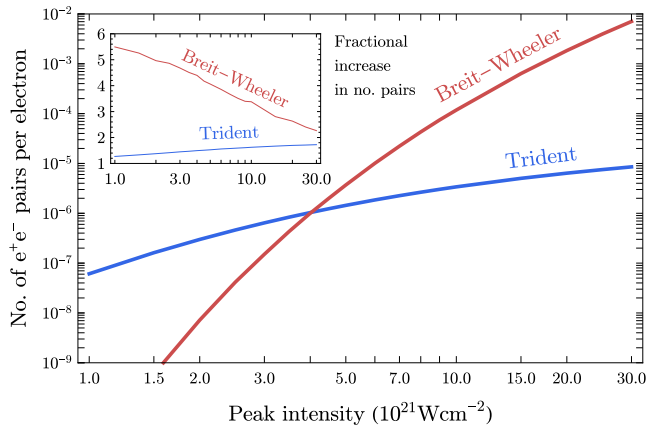}
	\caption{(Color online). The number of Breit-Wheeler (red) and trident
			(blue) electron-positron pairs produced per 1 GeV electron
			colliding with a linearly polarised plane-wave laser pulse with Gaussian
			temporal 	profile (FWHM 30 fs) and given intensity.
			{\it Inset} The fractional increase in the number of pairs for
			the same processes produced by treating radiation reaction
			probabilistically.}
	\label{fig:pairs1}
	\end{figure}

Our code also models pair production by both the gamma rays and the electrons.
Shown in Fig. \ref{fig:pairs1} are the number of pairs produced per electron
by a monoenergetic 1 GeV beam incident on a plane-wave laser pulse with given
peak intensity.

Pair production will form a key part of the new physics probed by future
high-intensity laser experiments. However, pairs do not become energetically
significant in the experimental setup we describe here because they are not
accelerated by the laser fields; thus their
probability of emitting a photon or creating an additional pair before exiting
the laser pulse is very small. Converting laser energy to pairs requires
much more intense, counterpropagating laser beams that can accelerate the
electron up to high energy between emission events \cite{Ner, Rol, Arka}.

The large increase in the number of Breit-Wheeler pairs shown in the main
part of Fig.
\ref{fig:pairs1} arises because the rate $\df \tau_\text{BW}/\df t$
is highly non-linear in $\chi$ and has maximum rate of growth for
$\chi = 0.62$. Reaching this switch-on with current
high-intensity lasers is strongly dependent on the yield of close to GeV
photons. As straggling enhances this yield, it also increases the yield of
pairs (see inset of Fig. \ref{fig:pairs1}). The number of trident pairs is
similarly increased, as straggling allows more high-energy electrons to
penetrate to the region of highest field intensity.

	\begin{figure}
	\includegraphics[width=\linewidth]{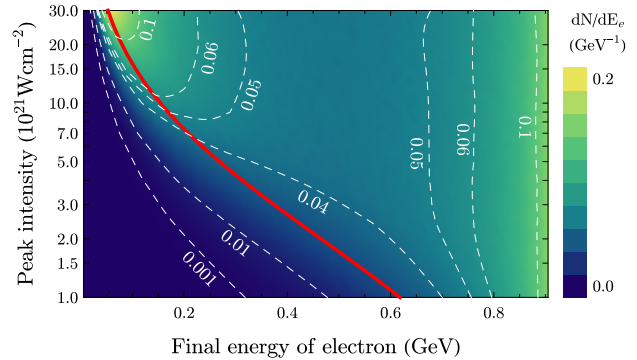}
	\caption{(Color online). The final state energy distribution
			$\text{d}N/\text{d}E_e$ of a 10 $\mu$m radius beam of 1 GeV
			electrons incident on a pulse with given peak intensity.
			White, dashed lines are contours of constant
			$\text{d}N/\text{d}E_e$. The red line is the lowest energy
			that can be reached with classical radiation.}
	\label{fig:electrons1}
	\end{figure}

Although potentially more difficult to measure, the electron energy
distribution can provide evidence of quantum radiation
reaction \cite{Neitz}. In Fig. \ref{fig:electrons1}, we show how the initally
monoenergetic beam acquires a spread 
because the electrons counterpropagate with a range of displacements from
the optical axis, experiencing different laser intensities. The energy loss
of electrons subject to continuous radiation reaction is determined only by
laser intensity. Thus there is a non-zero lower bound to the classical final
state energy distribution, corresponding to those electrons which have
passed through the region of highest laser intensity.

Discontinuously radiating electrons lose energy probabilistically, so some
electrons will straggle and lose more energy than possible classically.
At $10^{22} \, \text{Wcm}^{-2}$, $6.1 \times 10^{-3}$ of the beam electrons
will experience an energy loss exceeding 863 MeV, the maximum possible
classically. It may be possible to distinguish these electrons if the
initial energy of the beam is very well-characterised and devoid of a low
energy tail. This could be accomplished with a particle accelerator
\cite{SLAC}, or with magnetic filtering of the electron beam produced by a
laser-driven wakefield.

In this letter we have considered the effects of including a fully stochastic
model of radiation reaction on the motion of an energetic electron beam
incident on an intense laser pulse. Under the chosen conditions, the
parameter $\eta = 2 \gamma E_L/E_\mathrm{Sch} \sim 0.1$ and so QED effects
must be included. We find that electron motion is dominated by quantum
radiation reaction, enhancing the yield of photons with $\hbar \omega > 700$
MeV by a factor of $160$; thus the collision of a GeV electron beam with a
petawatt laser can provide evidence of both stochastic gamma ray and pair
production. The observation of 700 MeV photons will be an unambiguous
signature that GeV electrons have been incident on, and straggled through,
the region of highest intensity at the laser focus. The measurement of
increased energy spreading of the electron beam is likely to be more
ambiguous in a realistic experiment where laser parameters vary from
shot to shot. Experimental validation of the probabilistic nature of
QED-plasma models will underpin the simulation
and design of the next generation of higher energy laser-plasma interactions.

The authors thank N. Neitz for useful discussions. This work was
supported by an EPSRC studentship and funded in part by EPSRC grant
EP/G055165/1.


\begin{thebibliography}{99}
\bibitem{DiP}
	A. Di Piazza {\it et al.}, Rev. Mod. Phys. {\bf 84}, 1177 (2012)
\bibitem{DiP2}
	A. Di Piazza, K. Z. Hatsagortsyan and C. H. Keitel, Phys. Rev. Lett.
	{\bf 102}, 254802 (2009)
\bibitem{Elkina}
	N. V. Elkina {\it et al.}, Phys. Rev. ST Accel. Beams {\bf 14},
	054401 (2011)
\bibitem{Ner}
	E. N. Nerush {\it et al.}, Phys. Rev. Lett. {\bf 106}, 035001 (2011)
\bibitem{Ast}
	P. Goldreich and W. H. Julian, Astrophys. J. {\bf 157}, 869 (1969)
\bibitem{Tim}
	A. N. Timokhin, Mon. Not. R. Astron. Soc. {\bf 408}, 2092 (2010)
\bibitem{Ridgers}
	C. P. Ridgers {\it et al.}, Phys. Rev. Lett. {\bf 108}, 165006 (2012)
\bibitem{Bell}
	A. R. Bell and J. G. Kirk, Phys. Rev. Lett. {\bf 101}, 200403 (2008)
\bibitem{Sch}
	J. Schwinger, Phys. Rev. {\bf 82}, 664 (1951)
\bibitem{Sok}
	I. V. Sokolov {\it et al.}, Phys. Rev. Lett. {\bf 105}, 195005 (2010)
\bibitem{Ritus}
	V. I. Ritus, {\it Quantum effects of the interaction of elementary
	particles with an intense electromagnetic field}, Moscow Izdatel Nauka AN
	SSR Fizicheskii Institut Trudy {\bf 111}, 5 (1978)
\bibitem{Ugg2}
	U. I. Uggerh{\o}j, Rev. Mod. Phys. {\bf 7}, 1131 (2005)
\bibitem{Erber}
	T. Erber, Rev. Mod. Phys. {\bf 38} 4, 626 (1966)
\bibitem{Baier}
	V. N. Baier, V. M. Katkov and V. M. Strakhovenko, {\it Electromagnetic
	Processes at High Energies in Oriented Single Crystals},
	(World Scientific, Singapore, 1998)
\bibitem{Rol}
	R. Duclous, J. G. Kirk and A. R. Bell, Plasma Phys. Control. Fusion {\bf 53},
	015009 (2011)
\bibitem{Shen}
	C.S. Shen and D. White, Phys. Rev. Lett. {\bf 28} 7, 455 (1972)
\bibitem{Ugg}
	K. K. Andersen {\it et al.}, Phys. Rev. D {\bf 86}, 072001 (2012)
\bibitem{SLAC}
	D. L. Burke {\it et al.}, Phys. Rev. Lett. {\bf 79}, 1626 (1997)
\bibitem{Kneip}
	S. Kneip {\it et al.}, Phys. Rev. Lett. {\bf 103}, 035002 (2009)
\bibitem{Kneip2}
	S. Kneip {\it et al.}, Plasma Phys. Control. Fusion {\bf 53}, 014008 (2011)
\bibitem{Kim}
	H. T. Kim {\it et al.}, Phys. Rev. Lett {\bf 111}, 165002 (2013)
\bibitem{JCP}
	C. P. Ridgers {\it et al.}, arXiv:1311.5551 [physics.plasm-ph]
\bibitem{Furry}
	W. H. Furry, Phys. Rev. {\bf 81}, 115 (1951)
\bibitem{Brown}
	L. S. Brown and T. W. B. Kibble, Phys. Rev. {\bf 133}, A705 (1964)
\bibitem{Heinzl}
	T. Heinzl, Int. J. Mod. Phys. A27, 1260010 (2012)
\bibitem{Mack2}
	F. Mackenroth and A. Di Piazza, Phys. Rev. Lett. {\bf 110}, 070402 (2013)
\bibitem{Arka}
	J. G. Kirk, A. R. Bell and I. Arka, Plasma Phys. Control. Fusion {\bf 51},
	085008 (2009)
\bibitem{Mack}
	F. Mackenroth and A. Di Piazza, Phys. Rev. A {\bf 83}, 032106 (2011)
\bibitem{Seipt}
	D. Seipt and B. K\"{a}mpfer, Phys. Rev. A {\bf 83}, 022101 (2011)
\bibitem{LanLif}
	L. D. Landau and E. M. Lifshitz, {\it The Course of Theoretical Physics}
	Vol. 2, (Butterworth-Heinemann, Oxford, 1987)
\bibitem{Thomas}
	A. G. R. Thomas {\it et al.}, Phys. Rev. X {\bf 2}, 041004 (2012)
\bibitem{Neitz}
	N. Neitz and A. Di Piazza, Phys. Rev. Lett. {\bf 111}, 054802 (2013)

\end{thebibliography}
\end{document}